\pdfoutput=1
\documentclass[runningheads,a4paper]{llncs}

\usepackage{amssymb}
\usepackage{graphicx}
\usepackage{amsmath}
\usepackage{algorithmic}
\usepackage{algorithm}
\usepackage{url}
\urldef{\mailsa}\path|pat@dcs.gla.ac.uk|

\newlength{\halftextwidth}
\begin{document}

\mainmatter  

\title{Diamond-free Degree Sequences \\\
     TR-2010-318}

\titlerunning{Diamond-free Degree Sequences}

\author{Alice Miller and Patrick Prosser}

\authorrunning{Diamond-free Degree Sequences}

\institute{Computing Science,\\
Glasgow University, Glasgow, Scotland\\
\{alice/pat\}@dcs.gla.ac.uk \\}

\maketitle


\section{Introduction}
\label{sec:intro}
\vspace{-3mm}
We introduce a new problem, CSPLib problem number 50, to generate all degree sequences that have a
corresponding diamond-free graph with secondary properties. This problem
arises naturally from a problem in mathematics to do with balanced incomplete block designs; 
we devote a section of this paper to this. The problem itself is challenging with respect to 
computational effort arising from the large number of symmetries within the models. We introduce
two models for this problem. The second model is an improvement on the first, and this improvement
largely consists of breaking the problem into two stages, the first stage producing graphical degree sequences 
that satisfy arithmetic constraints and the second part testing that there exists a graph with 
that degree sequence that is diamond-free. We now present the problem in detail and then
give motivation for it. Two models are then presented, along with a listing of solutions. We then conclude and
suggest further work that might be done.

\section{Problem Definition}
\label{sec:probDefn}
\begin{figure}
\centering
\includegraphics[height=8.0cm,width=9.0cm]{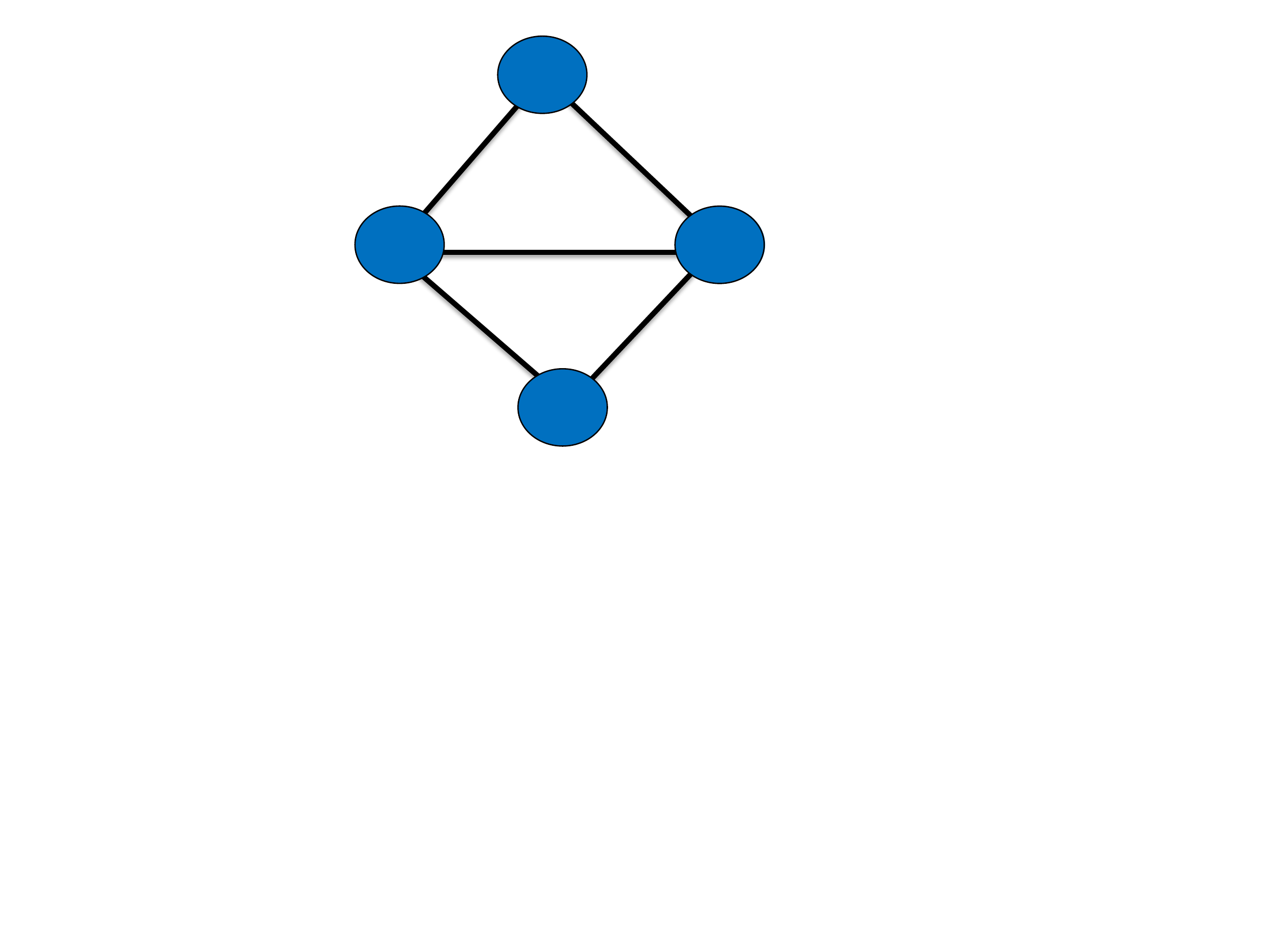}
\vspace{-4cm}
\caption{A simple diamond graph of four vertices and five edges.}
\label{fig:g1}
\end{figure}

\vspace{-3mm}
Given a simple undirected graph $G = (V,E)$, $V$ is the set of vertices and $E$ the set of undirected 
edges. The edge $\{u,v\} \in E$ if and only if vertex $u$ is adjacent to vertex $v$ in $G$. 
The graph is simple in that there are no loop edges, 
i.e. $\forall_{v \in V}~[\{v,v\} \notin E]$. Each vertex $v$ in $V$ has a degree 
$\delta(v) = |\{\{v,w\} : \{v,w\} \in E\}|$, i.e. the number of edges incident on that vertex.
A diamond is a set of four vertices in $V$ such that there
are at least five edges between those vertices (see Figure \ref{fig:g1} for an example of a diamond). Conversely, 
a graph is diamond-free if it has no diamond as a subgraph, i.e. for every set of four vertices 
the number of edges between those vertices is at most four. 

In our problem we have additional properties required of the degree sequences of the graphs, in particular
that the degree of each vertex is greater than zero (i.e. isolated vertices are disallowed), the degree
of each vertex is divisible by  3, and the sum of the degrees is divisible by  12 (i.e. $|E|$ is divisible by  6).

The problem is then for a given value of $n$, such that $|V| = n$, produce all unique degree sequences 
$\delta(1) \geq \delta(2) \geq ... \geq \delta(n)$ such that there exists a diamond-free graph with that degree sequence,
each degree is non-zero and divisible by  3, and the number of edges is divisible by  6.

\begin{figure}
\centering
\includegraphics[height=8.0cm,width=12.0cm]{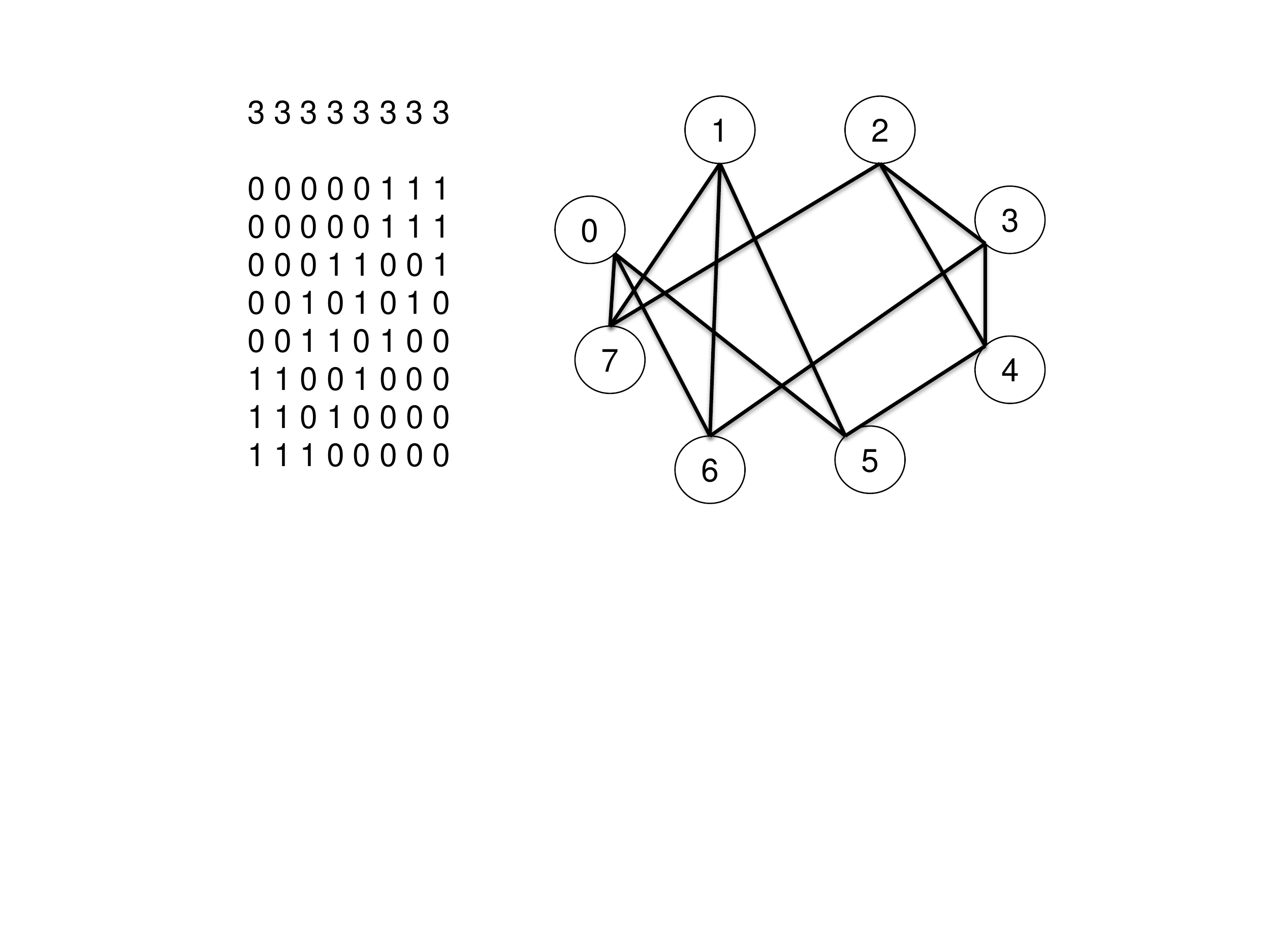}
\vspace{-4cm}
\caption{A degree sequence for $n = 8$ with the corresponding adjacency matrix and graph that is diamond-free.}
\label{fig:n8}
\end{figure}

\noindent
In Figure \ref{fig:n8} we give the unique degree sequence for $n=8$ and an adjacency matrix and simple graph that 
both corresponds to that sequence and represents a diamond-free graph.

\section{Motivation}
\label{sec:motivation}
\vspace{-3mm}
The problem is a byproduct of attempting to classify
partial linear spaces that can be produced during the execution of an
extension of Stinson's hillwalking algorithm for block designs with
block size $4$. First we need some definitions (see \cite{codihandbook}).
\begin{definition}
A Balanced Incomplete Block Design (BIBD) is a pair $(V,B)$ where $V$
is a set of $n$ points and $B$ a collection of subsets of $V$ (blocks)
such that each element of $V$ is contained in exactly $r$ blocks and
every $2$-subset 
of $V$ is contained in exactly $\lambda$ blocks.
\end{definition}

\noindent
Variations on $BIBD$s include {\it Pairwise Balanced Designs} (PBDs) in
which blocks can have different sizes, and {\it linear spaces} which
are PBDs in which every block has size at least $2$. It is usual to
refer to the blocks of a linear space as a {\it line}. A partial
linear space is a set of lines in which every pair appears in {\it at most}
$\lambda$ blocks. Here we refer to a BIBD with $\lambda=1$ as a {\it block design} and
to a partial linear space with $\lambda=1$, having $s_{i}$ lines of size $i$, where $i\geq 3$
and $s_{i}>0$ as a $3^{s_{3}}4^{s_{4}}\ldots$ structure.
For example, a block design on $7$ points with block size $3$ is given
by the following set of blocks: $$(0,1,2), (0,5,6), (0,3,4), (1,4,5),
(2,3,5), (2,6,4), (1,3,6)$$ and a $3^{4}4^{1}$ structure by the set of
blocks
$$(0,1,2,3), (0,4,5), (0,6,7), (1,4,6), (1,5,7)$$ Note that in the
latter case we do not list the lines of size $2$. 
Block designs with block size $3$ are known as Steiner Triple
Systems (STSs). These exist for all $n$ for which $n\equiv\;1,3\;({\rm
  mod}\;6)$ \cite{ki}. For example the block design given above is the
unique STS of order $7$ (STS($7$)). Similarly block designs with block
size $4$ always exist whenever $n\equiv\; 1,4\;({\rm mod}\;12)$.

\begin{algorithm}[t]
\caption{Algorithm to generate an STS on $n$ points}\label{alg:stinson3}
\begin{algorithmic}
\STATE $n\gets$ number of points 
\STATE LivePairs $\gets\{(i,j):0\leq i<j<n\}$ 
\STATE Blocks $\gets$ empty set
\WHILE{LivePairs not empty}
\STATE choose $(x,y)$ and $(y,z)$ from LivePairs
\STATE remove $(x,y)$ and $(y,z)$ from LivePairs
\STATE add $(x,y,z)$ to Blocks
\IF{$(y,z)$ is in LivePairs}
\STATE remove $(y,z)$ from LivePairs
\ELSE
\STATE remove existing block containing $(y,z)$, $(w,y,z)$
\STATE add $(w,z)$ to LivePairs
\ENDIF
\ENDWHILE
\end{algorithmic}
\end{algorithm}

Algorithm ~\ref{alg:stinson3} allows us to generate an STS for
any $n$ and is due to Stinson \cite{krst1}.
This algorithm always works, i.e. it never fails to terminate due to
reaching  a point where
the STS is not created and there are no suitable pairs $(x,y)$ and $(x,z)$.

\begin{algorithm}[t]
\caption{Algorithm to generate a block design with block size $4$ on $n$ points}\label{alg:stinson4}
\begin{algorithmic}
\STATE $n\gets$ number of points 
\STATE LiveTriples $\gets\{(i,j,k):0\leq i<j<k<n\}$ 
\STATE Blocks $\gets$ empty set
\WHILE{LiveTriples not empty}
\STATE choose $(x,y,z)$ and $(x,y,w)$ from LiveTriples
\STATE remove $(x,y,z)$ and $(x,y,w)$ from LivePairs
\STATE add $(x,y,z,w)$ to Blocks
\IF{$(y,z,w)$ is in LiveTriples}
\STATE remove $(y,z,w)$ from LiveTriples
\ELSE
\STATE remove existing block containing $(y,z,w)$, $(u,y,z,w)$
\STATE add $(u,y,z)$ and $(u,z,w)$ to LiveTriples
\ENDIF
\ENDWHILE
\end{algorithmic}
\end{algorithm}

A natural extension to this algorithm, for the case where block size
is $4$, is proposed in Algorithm \ref{alg:stinson4}.
This algorithm does not always work. It is possible for the algorithm
to fail to terminate due to reaching a point where the block design is
not created and there are no suitable overlapping triples $(x,y,z)$
and $(x,y,w)$ in LiveTriples. For this reason, we replace the condition
on the while loop by

\vspace{.25cm}
\noindent
\texttt{while LiveTriples not empty and
  overlapping triples exist}. 

\vspace{.25cm}
\noindent
Now the algorithm terminates, but rather than always producing a block
design, either produces a block design, or a $4^{s_{4}}$ structure, for which the complement has no overlapping
triples. I.e. the complement graph is diamond-free.

When $n=13$ the algorithm either produces a block design, or a $4^{8}$
structure whose complement graph consists of  $4$ non-intersecting
triangles.

The next open problem therefore is for $n=16$. If the algorithm does
not produce a block design, what is the nature of the structure it
does produce? To do this, we need to classify the $4^{r_{4}}$
structures whose complement graph is diamond-free.

The cases for which the $4^{s_{4}}$
structure has at least $2$ points that are in the maximum number of
blocks ($5$) are fairly straightforward. (There are fewer cases as
this number increases.). However if the number of such points is $0$ or
$1$, there is a large number of sub-cases to consider. The problem is
simplified if we can dismiss potential $4^{s_{4}}$
structures because the degree sequences of their complements can not be
associated with a diamond-free graph. This leads us to the problem outlined in this report: to
classify the degree sequences of diamond-free graphs of order $15$ and
$16$.
Note that each point that is not in $5$ blocks is either in no
blocks or is in blocks with in some number of points, where that
number is divisible by $3$. Thus for every point there is a vertex 
in the complement graph whose degree is also divisible by
$3$. In addition, since the number of pairs in both a block design on
$16$ points and a 
$4^{s_{4}}$ structure are divisible by $6$, the number of edges in the
complement  graph must be divisible by $6$.

\section{Constraint Models for Diamond-free Degree Sequences}
\label{sec:models}
\vspace{-3mm}
We present two constraint models for the diamond-free degree sequence problem. The first model we call
model A, the second model B. In many respects the two models are very similar but what is different
is how we solve them. In the subsequent descriptions we assume that we
have as input the integer $n$, where $|V| = n$ and vertex $i \in V$. All the constraint models were
implemented using the choco toolkit \cite{JChoco}. Further we assume that a variable $x$ has a domain of values
$dom(x)$.

\subsection{Model A}
\vspace{-3mm}
Model A is based on the adjacency matrix model of a graph.
We have a 0/1 constrained integer variable for each edge in the graph such that 
$A_{ij} = 1 \iff \{i,j\} \in E$.
In addition we have constrained integer variables $deg_1$ to $deg_n$ corresponding to the
degrees of each vertex, such that
\begin{eqnarray}
\forall_{i \in [1..n]} ~ dom(deg_{i}) = [3~..~n-1]
\end{eqnarray}
We then have constraints to ensure that the graph is simple:
\begin{eqnarray}
\forall_{i \in [1..n]} \forall_{j \in [i .. n]} ~ [A_{i,j} = A_{j,i}] \\
\forall_{i \in [1..n]} [A_{i,i} = 0]
\end{eqnarray}
Constraints are then required to ensure that the graph is diamond-free:
\begin{eqnarray}
\forall_{\{i,j,k,l\} \in V} [A_{i,j} + A_{i,k} + A_{i,l} + A_{j,k} + A_{j,l} + A_{k,l} \leq 4]
\end{eqnarray}
Finally we have constraints on the degree sequence:
\begin{eqnarray}
\forall_{i \in [1~..~n]}[deg_i = \sum_{j=1}^{j=n}A_{i,j}] \\
\forall_{i \in [1~..~n-1]}[deg_i \geq deg_{i+1}] \\
\forall_{i \in [1~..~n]} [deg_{i}~{\bf mod}~3 = 0] \\
\sigma = \sum_{i=1}^{i=n} deg_{i} \\
\sigma~{\bf mod}~12 = 0 
\end{eqnarray}
The vertex degree variables $deg_1$ to $deg_n$ are the decision variables.

\subsection{Model B}
\vspace{-3mm}
Model B is essentially model A broken into two parts, each part solved separately. The first part
of the problem is to produce a graphical degree sequence that meets the arithmetic constraints. The second part
is to determine if there exists a diamond-free graph with that degree sequence. Therefore solving proceeds
as follows.

\begin{enumerate}
\item Generate the next degree sequence $\pi = d_{1},d_{2},...,d_{n}$ that meets the arithmetical constraints.
If no more degree sequences exist then terminate the process.
\item If the degree sequence $\pi$ is not graphical return to step 1.
\item Determine if there is a diamond-free graph with the degree sequence $\pi$. 
\item Return to step 1.
\end{enumerate}

The first part of model B is is then as follows. Integer variables $deg_1$ to $deg_n$ correspond to the
degrees of each vertex and we satisfy constraints (1), (6), (7), (8) and (9) to generate a degree sequence.

Each valid degree sequence produced is then tested to determine if it is graphical using the 
Havel-Hakimi algorithm \cite{hh}. If the degree sequence is graphical we create an adjacency matrix as in
(2) and (3) and post the constraints (4) and (5) (diamond free with given degree sequence) where the variables $deg_{1}$ to $deg_{n}$ 
have been instantiated.
Finally we are in a position to post static symmetry breaking constraints. If we are producing a graph
and $deg_{i} = deg_{j}$ then these two vertices are interchangeable. Consequently we can insist that row $i$ in
the adjacency matrix is lexicographically less than or equal to row $j$. Therefore we post the following constraints:
\begin{eqnarray}
\forall_{i \in [1~..~n-1]}[deg_{i} = deg_{i+1} \Rightarrow A_{i} \preceq A_{i+1}]
\end{eqnarray}
where $\preceq$ means lexicographically less than or equal. In this second stage of solving the variables 
$A_{1,1}$ to $A_{n,n}$ are the decision variables.

\section{Solutions}
\label{sec:solutions}
\vspace{-3mm}
Our results are tabulated in Table \ref{tab1} (at end of report) for $8 \leq n \leq 16$. 
All our results are produced using model B run on a machine with 8 Intel Zeon E5420 processors 
running at 2.50 GHz, 32Gb of RAM, with version 5.2 of linux. The longest run time was for $n = 16$ taking about 5
minutes cpu time. Included in Table \ref{tab1} is the cpu time in seconds to generate all degree sequences for a given value of $n$.

All our results were verified. For each degree sequence the corresponding adjacency matrix 
was saved to file and verified to correspond to a simple diamond-free graph that matched the degree sequence and 
satisfied the arithmetic constraints. The verification software did not use any of the constraint programming code.

\section{Conclusion}
\label{sec:conc}
\vspace{-3mm}
We have presented a new problem, the generation of all degree sequences for diamond free
graphs subject to arithmetic constraints. Two models have been presented, A and B. Model A is
impractical, whereas model B is two stage and allows static symmetry breaking.

There are two possible improvements. The first is to model A. We might add the lexicographical
constraints, as used in model B, conditionally during search. The second improvement worthy
of investigation is to employ a mixed integer programming solver for the second stage 
of model B.

We are currently using the lists of feasible degree sequences for
diamond-free graphs with $15$ or $16$ vertices to simplify our proofs
for the classification of $4^{s_{4}}$ structures with diamond-free
complements, when the number of points in the maximum number of blocks
is $1$ or $0$ respectively. The degree sequence results for a smaller
number of points will also help to simplify our existing proofs for
cases where more points are in the maximum number of
blocks. Ultimately we would like to use our classification to modify
the extension of Stinson's algorithm for block size $4$ to ensure that
a block design is always produced. 

In the more distant future, we would like to analyse the structures
produced using our algorithm when $n>16$. The next case is $n=25$ and
the corresponding diamond-free graphs would have up to $25$
vertices.

\section*{Acknowledgments}
\label{sec:ack}
\vspace{-3mm}
We would like to thank Ian Miguel for helping us make CSPLib entry number 50, and Mike Codish and Brendan McKay for spotting 
errors in an earlier version of this report.

\bibliographystyle{plain}

\begin{table}
\begin{center}
\begin{tabular}{|cc|l|} \hline 
  $n$  & time & degree sequence \\ \hline 
 8 & 0.1 & 3 3 3 3 3 3 3 3 \\ \hline
 9 & 0.1 & 6 6 6 3 3 3 3 3 3 \\ \hline
10 & 0.5 & 6 6 3 3 3 3 3 3 3 3 \\ \hline
11 & 0.8 & 6 3 3 3 3 3 3 3 3 3 3 \\ \hline
12 & 1.4 & 3 3 3 3 3 3 3 3 3 3 3 3 \\
   & & 6 6 6 6 3 3 3 3 3 3 3 3 \\
   & & 6 6 6 6 6 6 6 6 6 6 6 6  \\
   & & 9 6 6 3 3 3 3 3 3 3 3 3  \\ \hline
13 & 3.7 & 6 6 6 3 3 3 3 3 3 3 3 3 3  \\
   & & 6 6 6 6 6 6 6 3 3 3 3 3 3  \\
   & & 6 6 6 6 6 6 6 6 6 6 6 3 3  \\
   & & 9 6 3 3 3 3 3 3 3 3 3 3 3  \\ \hline
14 & 14.0 & 6 6 3 3 3 3 3 3 3 3 3 3 3 3 \\ 
   & & 6 6 6 6 6 6 3 3 3 3 3 3 3 3 \\
   & & 6 6 6 6 6 6 6 6 6 6 3 3 3 3 \\
   & & 6 6 6 6 6 6 6 6 6 6 6 6 6 6 \\
   & & 9 3 3 3 3 3 3 3 3 3 3 3 3 3 \\
   & & 9 6 6 6 6 3 3 3 3 3 3 3 3 3 \\
   & & 9 9 6 6 3 3 3 3 3 3 3 3 3 3 \\
   & & 9 9 9 3 3 3 3 3 3 3 3 3 3 3 \\ \hline
15 & 107.7 & 6 3 3 3 3 3 3 3 3 3 3 3 3 3 3 \\
   & & 6 6 6 6 6 3 3 3 3 3 3 3 3 3 3 \\
   & & 6 6 6 6 6 6 6 6 6 3 3 3 3 3 3 \\
   & & 6 6 6 6 6 6 6 6 6 6 6 6 6 3 3 \\
   & & 9 6 6 6 3 3 3 3 3 3 3 3 3 3 3 \\
   & & 9 6 6 6 6 6 6 6 3 3 3 3 3 3 3 \\
   & & 9 6 6 6 6 6 6 6 6 6 6 6 3 3 3 \\
   & & 9 9 6 3 3 3 3 3 3 3 3 3 3 3 3 \\
   & & 9 9 6 6 6 6 6 3 3 3 3 3 3 3 3 \\
   & & 9 9 6 6 6 6 6 6 6 6 6 3 3 3 3 \\
   & & 9 9 9 6 6 6 3 3 3 3 3 3 3 3 3 \\
   & & 9 9 9 9 9 9 6 6 6 6 6 6 6 6 6 \\
   & & 12 6 6 3 3 3 3 3 3 3 3 3 3 3 3 \\
   & & 12 12 12 3 3 3 3 3 3 3 3 3 3 3 3 \\ \hline
16 & 339.8 & 3 3 3 3 3 3 3 3 3 3 3 3 3 3 3 3 \\
   & & 6 6 6 6 3 3 3 3 3 3 3 3 3 3 3 3 \\
   & & 6 6 6 6 6 6 6 6 3 3 3 3 3 3 3 3 \\
   & & 6 6 6 6 6 6 6 6 6 6 6 6 3 3 3 3 \\
   & & 6 6 6 6 6 6 6 6 6 6 6 6 6 6 6 6 \\
   & & 9 6 6 3 3 3 3 3 3 3 3 3 3 3 3 3 \\
   & & 9 6 6 6 6 6 6 3 3 3 3 3 3 3 3 3 \\
   & & 9 6 6 6 6 6 6 6 6 6 6 3 3 3 3 3 \\
   & & 9 9 3 3 3 3 3 3 3 3 3 3 3 3 3 3 \\
   & & 9 9 6 6 6 6 3 3 3 3 3 3 3 3 3 3 \\
   & & 9 9 6 6 6 6 6 6 6 6 3 3 3 3 3 3 \\
   & & 9 9 6 6 6 6 6 6 6 6 6 6 6 6 3 3 \\
   & & 9 9 9 6 6 3 3 3 3 3 3 3 3 3 3 3 \\
   & & 9 9 9 6 6 6 6 6 6 6 6 6 6 3 3 3 \\
   & & 9 9 9 9 3 3 3 3 3 3 3 3 3 3 3 3 \\
   & & 9 9 9 9 6 6 6 6 6 6 6 6 3 3 3 3 \\
   & & 9 9 9 9 6 6 6 6 6 6 6 6 6 6 6 6 \\
   & & 9 9 9 9 9 6 6 6 6 6 6 6 6 6 6 3 \\
   & & 9 9 9 9 9 9 6 6 6 6 6 6 6 6 3 3 \\
   & & 12 6 3 3 3 3 3 3 3 3 3 3 3 3 3 3 \\
   & & 12 9 9 6 3 3 3 3 3 3 3 3 3 3 3 3 \\
   & & 12 12 6 6 3 3 3 3 3 3 3 3 3 3 3 3 \\
   & & 12 12 9 3 3 3 3 3 3 3 3 3 3 3 3 3 \\ \hline
\end{tabular}
\end{center}
\caption{Degree sequences, of length $n$, that meet the arithmetic constraints and have a simple diamond-free graph. Tabulated
is $n$, cpu time in seconds to generate all sequences of length $n$ and those sequences.}
\label{tab1}
\end{table}

\end{document}